\documentstyle[12pt,epsf]{article}
\begin{document}
\begin{center}
{\Large {\bf MESONIC AND BINDING CONTRIBUTIONS TO THE $EMC$ EFFECT
IN A RELATIVISTIC MANY-BODY APPROACH}}
\end{center}

\vspace{0.5cm}

\begin{center}
{\Large {E. Marco$^1$, E. Oset$^1$ and P. Fern\'andez de
C\'ordoba$^{2}$  }}
\end{center}

\vspace{0.3cm}

{\small {\it
$^1$ Departamento de F\'{\i}sica Te\'orica and IFIC, Centro Mixto Universidad
de Valencia - CSIC, 46100 Burjassot (Valencia) Spain. \\

$^2$  Departamento de Matem\'atica Aplicada, Universidad Polit\'ecnica
de Valencia, Valencia, Spain.}}

\vspace{0.7cm}

\begin{abstract}
{\small{ Starting from a covariant relativistic formalism which uses 
nucleon and pion propagators in a nuclear medium, we revise the 
traditional effects of Fermi motion, binding and mesonic corrections to
the $EMC$ effect. The calculations are done in a very accurate way,
using precise nucleon spectral functions and meson self-energies and a good 
reproduction of the $EMC$ data is obtained outside the shadowing region which
is not explored.}}
\end{abstract}

\section{Introduction}

The $EMC$  effect is one of the most studied and debated
processes in the interface of particle and nuclear physics. It thus looks 
strange that one comes out with new ideas. Actually the ideas used here 
are not so new, they are those that with the pass of time have survived
a thorough scrutiny. Yet, as I will show here, all these ideas
were never put together and were not accounted for
at the level of precision that the present work has accomplished.

These ideas are the following:

1) Pionic effects, which are relevant around $x = 0.1 - 0.3$ \cite{1}.

2) Fermi motion, of relevance at $x > 0.6$ \cite{2}.

3) Binding effects, responsible for the dip around $x \simeq 0.5 - 0.6$
\cite{3}.

4) Correlation between energy and momentum accounted for by means
of nuclear spectral functions \cite{4}.

5) Relativistic effects \cite{5}.

In connection with these ideas let us see the novelties that our work
has contributed \cite{6}.

1) We use a relativistic formalism from the beginning. The
structure functions are written in terms of nucleon and pion propagators 
in the nuclear medium and hence we have a covariant relativistic
framework. One of the motivations to do so was to avoid the use of the 
flux factor introduced in Ref.~\cite{7} in order to account for
relativistic effects in the nonrelativistic calculations. Actually one
of our findings is that this prescription is numerically rather inaccurate
and should not be used as a substitute of a proper relativistic 
calculation.

2) Our approach leads automatically to the formalism of spectral functions
used in \cite{4}, only that it appears in a relativistic form.

3) Mesonic effects have been reevaluated. Improvements appear in three
fronts.

a)  Static approximations relating mesonic effects to the ``pion
excess'' are proved to be inaccurate and  they are avoided.

b) The results are written in terms of the pion propagator $D (q)$ in
the medium. We prove that it is essential that the pion propagator
fulfills the sum rule

\begin{equation}
\int^\infty_0 \frac{dp^0}{\pi} (-) Im D (p^0, p) 2p^0 = 1\,,
\end{equation}

\noindent
which is the statement of the equal time commutation of the pion fields
expressed in momentum space.

The satisfaction of Eq.~(1) requires exact analytical properties of
the pion propagator, which most approximations used in the Literature do not
fulfill.

c) We included the effects of the $\rho$-meson cloud for the first time.

\section{Formalism}

We evaluate the self-energy for the electron corresponding to the
diagram of Fig.~1 in infinite nuclear matter.

\medskip
\begin{figure}
\hspace{1.0in}
\epsfxsize=3.0in
\centerline{\epsffile{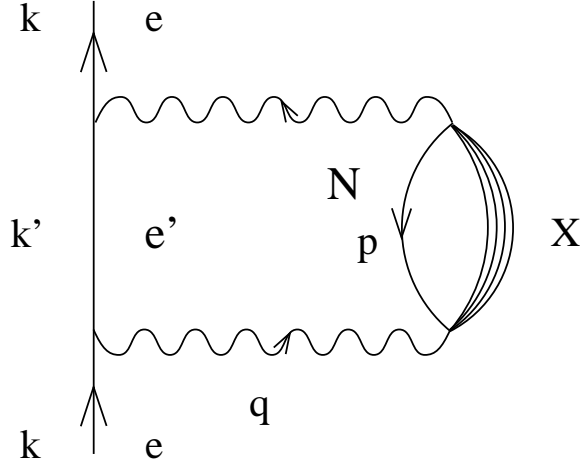}}
\caption{Electron self-energy associated with the process of
deep inelastic electron-nucleon scattering.
}
\label{fig1}
\end{figure}
\medskip

The probability per unit time of the electron interacting with the
nucleons in the medium to give the final state is given by

\begin{equation}
\Gamma (k) = - \frac{2m}{E_e (\vec{k}\,)} Im \Sigma (k)
\end{equation}

\noindent
and this can be reconverted into the contribution to the cross section
by an element of nuclear volume by means of

$$
d \sigma = \Gamma dtdS = \Gamma \frac{dt}{dl} dldS = \frac{\Gamma}
{v} d^3 r=
$$

\begin{equation}
= \Gamma \frac{E_e (\vec{k}\,)}{k} d^3 r = - \frac{2m}{k} Im \Sigma d^3 r\,.
\end{equation}
Hence the cross section for $eA$ scattering is given by

\begin{equation}
\sigma = \int d^3 r (\frac{-2m}{k}) Im \Sigma (\rho (r)) d^3 r\,,
\end{equation}

\noindent
where we have used the fact that $\Sigma$ is a function of the nuclear 
density and we substitute $\rho$ by $\rho (\vec{r})$, the experimental
density of the nucleus under consideration. So we are led in a natural
way to the local density approximation (LDA) which is a
highly accurate tool for this purpose. Indeed, we have also evaluated
the structure function using the finite nucleus
spectral function of Ref.~\cite{8} and found differences of the
order of 2 $\%$ in the EMC region with respect
to the LDA calculation using the same input.

The relativistic nucleon propagator in a medium is given by

\begin{equation}
G (p^0, p) = \frac{1}{p\!\!/ - M - \Sigma }
\end{equation}

\noindent
where 

\begin{equation}
\Sigma (p) \equiv \Sigma^s + \Sigma^v \gamma^0 + \Sigma^{v'}
\vec{\gamma} \vec{p}
\end{equation}
We separate the nucleon propagator into the positive energy and
negative energy parts as

$$
G (p^0, p) = \frac{M}{E (\vec{p})\,} \left\{ \sum_r u_r (\vec{p}\,)
\bar{u}_r (\vec{p}\,) \left[ \frac{1- n (\vec{p}\,)}{p^0 - E (\vec{p}\,)
+ i \epsilon} + \frac{n (\vec{p}\,)}{p^0 - E (\vec{p}\,) - i \epsilon} 
\right]
\right.
$$

\begin{equation}
\left. + \frac{\sum_r v_r (- \vec{p}\,) \bar{v}_r (- \vec{p}\,)}{p^0
 + E (\vec{p}\,) - i \epsilon } \right\}\,.
\end{equation}
We omit then the negative energy components and make an expansion of 
$(p\!\!/ - M - \Sigma )^{-1}$ in terms of the positive energy part
of the propagator of Eq.~(7). By doing this, one benefits from the fact
that all the terms of the nucleon self-energy of Eq.~(6) are diagonal in the
spinors $u_r$. For instance $\bar{u}_r (\vec{p}\,) \gamma^0 u_s (\vec{p}\,)
\propto \delta_{rs}$.

This allows one to write finally an expression for the nucleon
propagator in terms of relativistic spectral functions as

$$
G (p^0,p) = \frac{M}{E (\vec{p}\,)} \sum_r u_r (\vec{p}\,) \bar{u}_r
(\vec{p}\,) \left[ \int_{- \infty}^\mu d \omega \frac{S_h (\omega, p)}{
p^0 - \omega - i \eta} \right.
$$

\begin{equation}
\left. + \int_\mu^\infty d \omega \frac{S_p (\omega, p)}{p^0
- \omega + i \eta } \right]
\end{equation}

\noindent
with a normalization for $S_h (\omega,p)$ given by

\begin{equation}
4 \int d^3 r  \int \frac{d^3 p}{(2 \pi)^3} \int^\mu_{- \infty}
S_h (\omega, p, k_F (\vec{r}\,)) d \omega = A\,,
\end{equation}

\noindent
which is obtained by demanding that  the baryonic number is $A$ (this
was the
justification to introduce the flux factor in Ref.~\cite{7}).

In omitting the negative energy terms in the propagator we are
neglecting zigzag terms contributing to the nucleon self-energy. These,
and other possible terms missing in our approach to the nucleon
self-energy are recovered later on by adding a phenomenological piece to
the nucleon self-energy such that the binding energy obtained for each 
nucleus corresponds to the experimental one. Hence our emphasis is
in using input as consistent as possible with experimental information
in order to make the results highly accurate and as model independent
as possible.

\section{Deep inelastic cross section}

By explicitly evaluating Eq.~(4) using the nucleon propagator of
Eq.~(8) we find

\begin{equation}
\sigma_{A} = \frac{\alpha^2}{k} \int \frac{d^3 k'}{E_e (\vec{k}' \,)}
L'_{\mu \nu} W'^{\mu \nu}_A\,,
\end{equation}

\noindent
where $L'_{\mu \nu}$ is the leptonic tensor

\begin{equation}
L'_{\mu \nu} = 2 k_\mu k'_\nu + 2 k'_\mu k_\nu + q^2 g_{\mu \nu}
\end{equation}

\noindent
and $W'^{\mu \nu}$ the hadronic tensor given by

$$
\begin{array}{c}
W'^{\mu \nu} = 4 \int d^3 r \int \frac{d^3 p}{(2 \pi)^3} \frac{M}{E (\vec{p}
\,)} \int^\mu_{- \infty} d p^0 S_h (p^0, p)
$$

$$
W'^{\mu \nu} (p,q)
\end{array}
$$

\begin{equation}
p \equiv (p^0, \vec{p}\,) ; \; W^{\mu \nu} = \frac{1}{2}
(W'^{\mu \nu}_p + W'^{\mu \nu}_n)
\end{equation}

\noindent
for the nucleonic contribution to the hadronic tensor

Recalling gauge invariance $W'^{\mu \nu}$ can be written in terms
of two invariant structure functions.

\begin{equation}
W'^{\mu \nu} = \left( \frac{q^\mu q^\nu}{q^2} - g^{\mu \nu} \right) W_1 + 
\left( p^\mu - \frac{p . q}{q^2} q^\mu \right) \left( p^\nu -
\frac{p . q}{q^2} q^\nu \right) \frac{W_2}{M^2}
\end{equation}
In the Bjorken limit $(- q^2 \equiv Q^2 \rightarrow \infty, \;
q^0 \rightarrow \infty)$ we define

\begin{equation}
\begin{array}{ll}
x_N = \frac{-q^2}{2 p.q} \quad ; & \nu_N = \frac{p . q}{M}\\[2ex]
x = \frac{- q^2}{2Mq^0} \quad ; & \nu = \frac{M q^0}{M} = q^0
\end{array}
\end{equation}

\noindent
where the variables $x_N , \nu_N$ refer to the nucleons of the nuclear
medium and $x, \nu$ are the corresponding variables for nucleons at
rest in the rest nuclear frame.

One then introduces $F_1, F_2$, the Bjorken structure functions which in the
Bjorken limit depend only on $x$ and are defined as

\begin{equation}
\begin{array}{c}
\nu W_2 (x, Q^2) \equiv F_2 (x)\,, \\[2ex]
M W_1 (x, Q^2) \equiv F_1 (x)\,,
\end{array}
\end{equation}

\noindent
which furthermore satisfy the Callan-Gross relation

\begin{equation}
2 x F_1 (x) = F_2 (x)\,.
\end{equation}
Eq.~(12) in the Bjorken limit becomes

\begin{equation}
F_{2A,N} (x) = 4 \int d^3r \int \frac{d^3p}{(2 \pi)^3} \frac{M}{E (\vec{p}
\,)} \int^\mu_{- \infty} d p^0 S_h (p^0,p) \frac{x}{x_N} F_{2N} (x_N)
\theta  (x_N) \theta (1 - x_N)\,,
\end{equation}

\noindent
which gives us the nucleonic contribution to the nuclear structure function.

On coming to this point is worth commenting why nuclear effects
matter in the EMC effect. One often hears questions from our high 
energy colleagues of ``how can binding and Fermi motion effects matter
when one is performing experiments with electrons of hundreds of GeV?''.
The answer is shown in Eq.~(17) since the $F_{2N}$ structure function
appears with argument $x_N$ not $x$, and we can
prove that in the Bjorken limit

\begin{equation}
\frac{x}{x_N} = \frac{p^0 - p^3}{M}\,.
\end{equation}

\noindent
($\vec{q}$ has been chosen in the $z$ direction). So what Eq.~(18) shows is
that one is gauging the binding energies and the momenta against the 
nucleon mass, not against $Q^2$ or $q^0$.

For the mesonic contribution one evaluates the electron self-energy
corresponding to the diagram of Fig.~2,
from where we have to subtract the terms linear in the density since
they are already included in the nucleon  structure function 
evaluated in Eq.~(17).

\medskip
\begin{figure}[b]
\hspace{1.0in}
\epsfxsize=5.0in
\centerline{\epsffile{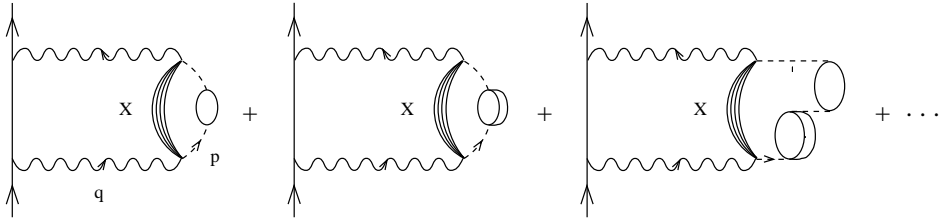}}
\caption{Diagrams of the electron self-energy including $1 ph, 
1 \Delta h, 1 ph 1 \Delta h$, etc.. 
}
\label{fig2}
\end{figure}
\medskip

Hence, one finds a pionic contribution, from the pion renormalization
in the  medium, additional to the nucleonic one calculated before,
given by

\begin{equation}
F_{2A,\pi} (x_A) = -6 \int d^3 r \int \frac{d^4 p}{(2 \pi)^4}
\theta (p^0\,) \delta Im D (p) \frac{x}{x_\pi} 2M F_{2\pi} (x_\pi)
\theta (x_\pi - x) \theta  (1 - x_\pi)
\end{equation}

\noindent
with

\begin{equation}
\frac{x}{x_\pi} = \frac{- p^0 + p^3}{M}
\end{equation}

\noindent
and

\begin{equation}
\delta D = D - D_0 - \left.\frac{\partial D}{\partial \rho}\right|_{\rho = 0} 
\; \rho  
\end{equation}

\noindent
and a similar expression for the $\rho$ meson contribution.
 
It is worth discussing briefly which are the approximations
implicitly made when one uses the approximation of the  ``pion
excess in nuclei''. One can visualize it from our approach. If we
neglect the fact that $F_{2N} (x_\pi )$ depends on $p^0$ and 
that one has the strict $\theta$ functions from phase space
requirements in Eq.~(19), we could integrate $Im \delta D$ and
have 

\begin{equation}
\frac{\delta N_\pi (\vec{p}\,)}{2 \omega (\vec{p}\,)} = - 3 \int_0^\infty
\frac{dp^0}{2 \pi} \delta Im D (p)
\end{equation}

\noindent
and then one would have $F_{2A, \pi}$ as an integral over $\vec{p}$ of the
``pion excess'' distribution in the nucleus. It is clear that
this ``approximation'' is forcing the contribution of Eq.~(19) for
values of $p^0$ forbidden by energy and momentum conservation. Numerically
it leads to an overestimate of the mesonic effects and should be avoided.

In Fig.~3 we show now the results for the ratio $R_N$ defined as

\begin{equation}
R_N (x) = \frac{F_{2A,N} (x_A)}{A F_{2N} (x)}\,.
\end{equation}

\medskip
\begin{figure}[hb]
\hspace{1.0in}
\epsfxsize=4in
\centerline{\epsffile{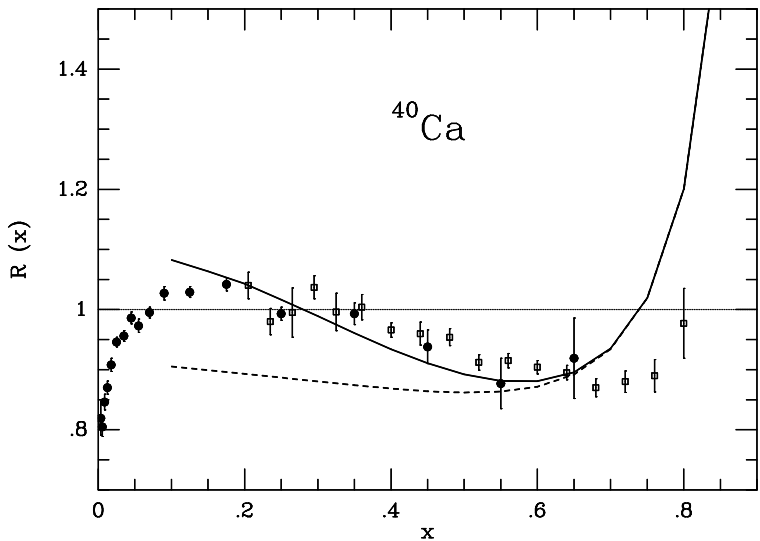}}
\caption{Results for $R(x)$ for $^{40}$Ca. Solid lines: whole
calculation including the nucleons and the mesons; dashed line: contribution
of the nucleons.
}
\label{fig3}
\end{figure}
\medskip

We divide by $F_{2N} (x)$, the nucleon structure function, instead
of dividing by the deuteron structure function as it corresponds
to the data. The reason is that we cannot use the LDA approximation
to evaluate the deuteron structure function.

Since the ratio of the deuteron structure function to twice the average
one of the nucleon is practically unity up to $x \simeq 0.6$ and then
becomes bigger than unity, we should expect an overestimate of the
experimental ratio $R_N$ in the 
$x > 0.6$ region, as it is the case. However, in Refs.~\cite{9,10},
where we evaluate the structure function for $x > 1$ we show 
absolute values which are in agreement with the experiment. The results for
$^{40}$Ca agree well with experiment outside the shadowing region
which we have not studied here. In Fig.~3 the dashed line is the nucleonic
contribution alone while the solid line includes also the mesonic
effects.

In Fig.~4 we show the same results for $^{56}$Fe but we show explicitly
the contribution of the pions (intermediate line) and the pions plus
$\rho$-meson, (upper line), (the lower one indicates the nucleonic
contribution alone).

We observe that the pionic contribution is more moderate than the one
found in Refs.~\cite{1} or \cite{11}. On the other hand our results
for the nucleonic contribution are about 10$\%$ lower around
$x = 0$ than those of Ref.~\cite{12}, where a nonrelativistic calculation
is done and the flux factor of Ref.~\cite{7} is used.

\medskip
\begin{figure}
\hspace{1.0in}
\epsfxsize=4.0in
\centerline{\epsffile{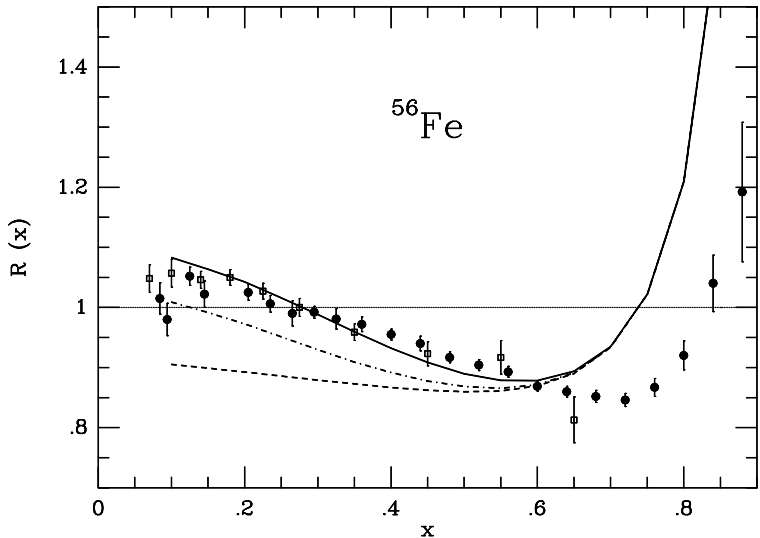}}
\caption{Results for $R(x)$ for $^{56}$Fe. Solid line: 
whole calculation including the nucleons and the mesons; dashed line:
contribution of the nucleons; dot-dashed line: contribution of nucleons 
plus pions.
}
\label{fig4}
\end{figure}
\medskip

\section{Conclusions}

We have performed a very accurate calculation of the nuclear
structure function in the region of the EMC effect
which incorporates the effects which have been found relevant
in the past: relativistic effects, binding, Fermi motion and
mesonic contributions.

We have put all these elements together for the first time
as well as the effects from the renormalization of the $\rho$-meson
cloud. 

When all these elements are put together we find a good agreement with
experiment outside the shadowing region which is not explored.

This suggests that the nucleon, resonances and mesonic degrees of freedom
are an adequate tool to deal with the many-body problem in
deep inelastic scattering. Furthermore, we showed that  the many-body
effects are quite relevant in spite of the large energies of the leptons
compared to the scale of the binding nuclear energies and Fermi 
momenta, and thus must be adequately considered in whichever other 
framework or other degrees of freedom  one chooses to study deep
inelastic lepton scattering in nuclei.

\end{document}